\documentclass[english,twocolumn,prl,showpacs,floatfix,nopac]{revtex4} 
\usepackage[T1]{fontenc}
\usepackage[latin9]{inputenc}
\usepackage{graphicx}
\usepackage{amssymb}
\usepackage{amsmath}%

\makeatletter

\usepackage{amsfonts}


\makeatother

\usepackage{babel}

\begin{document}

\title{Stochastic simulations of fermionic dynamics with phase-space representations}

\author{M. \"{O}gren, K. V. Kheruntsyan, J. F. Corney}

\affiliation{ARC Centre of Excellence for Quantum-Atom Optics, School of Mathematics
and Physics, University of Queensland, Brisbane, Queensland 4072,
Australia}

\date{\today{}}
\begin{abstract}
A Gaussian operator basis provides a means to formulate phase-space simulations
of the real- and imaginary-time evolution of quantum systems. Such simulations are guaranteed
to be exact while the underlying distribution remains well-bounded,
which defines a useful simulation time. We analyse the application
of the Gaussian phase-space representation to the dynamics of the
dissociation of an ultra-cold molecular gas. We show how the choice
of mapping to stochastic differential equations can be used to tailor
the stochastic behaviour, and thus the useful simulation time. In the phase-space approach, it is only averages of stochastic trajectories that have a direct physical meaning.   Whether particular constants of the motion are satisfied by individual trajectories depends on the choice of mapping, as we show in examples.
\end{abstract}


\maketitle




Numerical approaches are an indispensable
part of endeavours to understand quantum many-body physics in condensed
matter and AMO physics. In particular, there is a need for real-time,
dynamical simulations, driven in large part by the progress in the
control and flexibility of ultra-cold atom experiments, which has
made the dynamically evolving quantum many-body state more directly accessible.
For bosons, first-principles phase-space methods have successfully
simulated dynamics in experimentally realistic systems~\cite{DeuarPRL2007,SavagePRA2006}.
However, these methods are not directly applicable to fermionic systems,
which are an increasingly important area of ultra-cold atoms, often
with direct relevance to condensed matter systems.

The exponential growth of the Hilbert space with system size hinders
a brute-force approach for systems of more than a few modes.
Stochastic approaches,
provided they are unbiased, can provide exact results within the precision
determined by sampling error. A range of quantum Monte Carlo methods
has been used to address a variety of problems in many-body quantum
physics. However, the limitations when it comes to dynamics are well
known~\cite{Linden1992}, for example, the oscillating phase problem
in path-integral approaches~\cite{MakJCP2009}. An interesting direction
in recent years has been the extension to fermionic systems of stochastic
wavefunction approaches~\cite{MontinaPRA2006}.

In this work we employ a Gaussian stochastic method based on a generalized
phase-space representation of the quantum density operator~\cite{CorneyPRL2004}.
The representation allows the quantum Liouville equation for the density operator to be mapped onto an equivalent  Fokker-Planck equation for a distribution function over phase space, provided that the distribution vanishes at the boundary.  This distribution is then sampled via equivalent stochastic phase-space equations, with physical results corresponding to stochastic averages.   The phase-space equations are structurally
similar to the Heisenberg equations for the corresponding operators, with additional stochastic noise.

Here we explore the freedom in choosing the stochastic noise in order to reduce sampling errors and extend the useful simulation time.  We first introduce the Bose-Fermi model we use study these effects, and a set of conserved quantities that can be used to benchmark the different choices of stochastic equations.    After reviewing the phase-space formalism, we give the general form of the stochastic equations corresponding to the Hamiltonian.  To exemplify the gauge freedom,  we then give two forms of the noise terms and demonstrate through simulation their very different numerical properties.

As a particular application, we consider a model of production of correlated
pairs of fermionic atoms by dissociation of a Bose-Einstein condensate
(BEC) of diatomic molecules~\cite{JackPRA2005,KheruntsyanPRL2006}.
The uniform molecular BEC is
initially in a coherent state at zero temperature with average initial number of molecules $N_0$, with no atoms present.
The created fermionic atoms are modeled as being untrapped in the
x-direction, and propagate through the homogeneous condensate. The
Hamiltonian of this boson-fermion model~\cite{FriedbergPRB1989}
is given by \begin{equation}
\widehat{H}=\hbar\sum\nolimits _{\mathbf{k},\sigma}\Delta_{\mathbf{k}}\hat{n}_{\mathbf{k},\sigma}-i\hbar\kappa\sum\nolimits _{\mathbf{k}}\left(\hat{a}^{\dagger}\hat{m}_{\mathbf{k}}-\hat{m}_{\mathbf{k}}^{\dagger}\hat{a}\right),\label{Hamiltonian}\end{equation}
where $\mathbf{k}$ labels the $M$ plane-wave modes for a quantization box of length $L$ and $\sigma=1,2$
labels the effective spin state for the atoms. The fermionic number
and pair operators are defined as $\hat{n}_{\mathbf{k},\sigma}=\hat{c}_{\mathbf{k},\sigma}^{\dagger}\hat{c}_{\mathbf{k},\sigma}$
and $\hat{m}_{\mathbf{k}}=\hat{c}_{\mathbf{k},1}\hat{c}_{-\mathbf{k},2}$,
respectively, with $\{\hat{c}_{\mathbf{k},\sigma},\hat{c}_{\mathbf{k}^{\prime},\sigma^{\prime}}^{\dagger}\}=\delta_{\mathbf{k},\mathbf{k}^{\prime}}\delta_{\sigma,\sigma^{\prime}}$,
while the bosonic molecular operators obey $[\hat{a},\hat{a}^{\dagger}]=1$. The first term gives the kinetic energy of the atoms of mass $m_a$ and the detuning $\Delta$ between the atomic and molecular levels: $\hbar\Delta_{\mathbf{k}}\equiv\hbar^{2}\left\vert \mathbf{k}\right\vert ^{2}/(2m_{a})+\hbar\Delta$.
The second term describes the atom-molecule coupling of strength  $\kappa$ .

The physics of the growth of correlations during the dynamics has been explored elsewhere~\cite{Ogren2010}.  Here, we use the same system parameters but focus on the evolution of certain conserved quantities. While such quantities are constant in the stochastic averages, which have a physical meaning, they are not necessarily constant in individual trajectories.  We study this evolution to monitor the growth of sampling error for different choices of the stochastic equations, and to illustrate the exactness of the method within the limitations of sampling error.

The spin-symmetry of the Hamiltonian implies the identity $\hat{n}_{\mathbf{k}}\equiv \hat{n}_{\mathbf{k},1} =\hat{n}_{-\mathbf{k},1}= \hat{n}_{\mathbf{k},2}=\hat{n}_{-\mathbf{k},2}$
for equal initial populations.  An additional operator identity
arises from the homogeneity of the molecular condensate,  
\begin{equation}
\hat{m}_{\mathbf{k}}^{\dagger}\hat{m}_{\mathbf{k}}\left(=\hat{n}_{\mathbf{k},1}\hat{n}_{-\mathbf{k},2}\right)=\hat{n}_{\mathbf{k}}\label{MPMrelation}.
\end{equation}
According to this, we expect the conserved quantity
\begin{equation}
F_{\mathbf{k}} \equiv \left\langle \widehat{m}_{\mathbf{k}}^{\dagger}\widehat{m}_{\mathbf{k}}\right\rangle - \left\langle \widehat{n}_{\mathbf{k}}\right\rangle \label{MPMQaveragerelation}
\end{equation}
to be zero in any numerical implementation.
We calculate this quantity numerically for the resonant Fourier mode $\mathbf{k}_0$, along with the total energy normalised by the dissociation energy: $E \equiv \langle \hat H \rangle/2\hbar | \Delta |$ and the total number of molecules and pairs, normalised by the initial number molecules: $N \equiv ( 2\langle \hat{a}^\dagger \hat{a}\rangle + \sum_{\mathbf{k}, \sigma} \langle \hat n_{\mathbf{k}, \sigma} \rangle)/2N_0$, which is also conserved.
%


The Gaussian phase-space representation maps pairs of annihilation/creation
operators onto first-order differential operators. It can thereby be used to transform the Liouville equation for unitary evolution
\begin{equation}
\frac{d}{dt} \hat \rho = -\frac{i}{\hbar}\left[\hat H, \hat \rho \right] \label{LE}
\end{equation}
into a differential equation for an equivalent phase-space distribution, so long as certain boundary terms vanish.  In practice the appearance of a boundary term is indicated by the rapid growth of sampling error and the appearance of large excursions in the trajectories~\cite{GilchristPRA1997}, and this places a limitation on the length of the simulation.  For Hamiltonians containing up to four operators, a second-order partial differential equation is generated, which can be written in the form of a Fokker-Planck equation (FPE):
\begin{equation}
\frac{d}{dt} P(\vec \lambda) = \left[ - \sum_j \frac{\partial}{\partial \lambda_j} A_j(\vec \lambda)  +  \frac{1}{2} \sum_{j,k}\frac{\partial^2}{\partial \lambda_j\partial \lambda_k} D_{ij}(\vec \lambda) \right] P(\vec \lambda).\label{generalFPE}
\end{equation}
The first order derivatives in the phase-space variables $\lambda_j$ correspond to drift behaviour
and the second order to the diffusion.
Effects such as three-body interactions will result in higher-order derivatives, but these are difficult to efficiently sample by numerical methods \cite{PlimakEPL2001}.

In general the phase-space variables $\vec \lambda$ are complex. However, the analytic nature of the Gaussian operators gives a freedom in the choice of derivatives when the variables are expanded into real and imaginary parts.  The diffusion matrix $D$ of the resulting FPE can always be chosen to be positive-definite \cite{RahavPRB2009}, as required for stochastic sampling.

The quantum state generated by the Hamiltonian (\ref{Hamiltonian}) can be represented by a distribution over $3M+2$ variables $\vec{\lambda}\left(t\right)=\left(n_{1},\dots,n_{M},m_{1},\dots,m_{M},m_{1}^{+},\dots,m_{M}^{+},\beta,\beta^{+}\right)$,
with $m_{j}^{+}\neq m_{j}^{\ast}$ and $\beta^{+}\neq\beta^{\ast}$.  The corresponding FPE for the dynamics  is
\begin{equation}
\begin{array}{l}
\partial_{t}P= 2i\sum_{\mathbf{k}}\Delta_{\mathbf{k}}\left[\partial_{m_{\mathbf{k}}} m_{\mathbf{k}} -\partial_{m_{\mathbf{k}}^{+}} m_{\mathbf{k}}^{+}\right]P\\
+\kappa\sum_{\mathbf{k}}\left[-\partial_{n_{\mathbf{k}}}\left(\beta^{+}m_{\mathbf{k}}+\beta m_{\mathbf{k}}^{+}\right)-\partial_{m_{\mathbf{k}}}\beta\left( 1-2n_{\mathbf{k}} \right)\right.\\
-\partial_{m_{\mathbf{k}}^{+}}\beta^{+}\left(1-2n_{\mathbf{k}} \right)+\partial_{\beta} m_{\mathbf{k}} +\partial_{\beta^{+}} m_{\mathbf{k}}^{+} \\
+\partial_{n_{\mathbf{k}}}\partial_{\beta} n_{\mathbf{k}} m_{\mathbf{k}}+\partial_{n_{\mathbf{k}}}\partial_{\beta^{+}} n_{\mathbf{k}}m_{\mathbf{k}}^{+} +\partial_{m_{\mathbf{k}}} \partial_{\beta} m_{\mathbf{k}}^2 \\
\left.-\partial_{m_{\mathbf{k}}} \partial_{\beta^{+}}  n_{\mathbf{k}}^2 -\partial_{m_{\mathbf{k}}^{+}}\partial_{\beta} n_{\mathbf{k}}^2 +\partial_{m_{\mathbf{k}}^{+}}\partial_{\beta^{+}}  m_{\mathbf{k}}^{+2} \right] P.\end{array}\label{PDE}\end{equation}
Note that all differential operators act also on the multidimensional
distribution $P=P\left(\vec{\lambda},t\right)$.
To directly solve Eq.~(\ref{PDE}) is computationally unfeasible
for many variables.  Instead one can employ a mapping \cite{GardinerBook1, NumSim} to an equivalent set of stochastic differential
equations (SDEs) to sample the moments of the distribution.  In the Ito calculus, stochastic equations corresponding to Eq.~(\ref{PDE}) have the general form
\begin{equation}
\begin{array}{l}
dn_{\mathbf{k}}=\left(\alpha m_{\mathbf{k}}^{+}+\alpha^{+}m_{\mathbf{k}}\right)d\tau+N_{0}^{-1/2}\mathbf{B}^{\left(n_{\mathbf{k}}\right)}\mathbf{dW},\\
dm_{\mathbf{k}}=\left[-2i\delta_{\mathbf{k}}m_{\mathbf{k}}+\alpha\left(1-2n_{\mathbf{k}}\right)\right]d\tau+N_{0}^{-1/2}\mathbf{B}^{\left(m_{\mathbf{k}}\right)}\mathbf{dW},\\
dm_{\mathbf{k}}^{+}=\left[2i\delta_{\mathbf{k}}m_{\mathbf{k}}^{+}+\alpha^{+}\left(1-2n_{\mathbf{k}}\right)\right]d\tau+N_{0}^{-1/2}\mathbf{B}^{\left(m_{\mathbf{k}}^{+}\right)}\mathbf{dW},\\
d\alpha=-\frac{1}{N_{0}}\sum_{\mathbf{k}}m_{\mathbf{k}}d\tau+N_{0}^{-1/2}\mathbf{B}^{\left(\alpha\right)}\mathbf{dW},\\
d\alpha^{+}=-\frac{1}{N_{0}}\sum_{\mathbf{k}}m_{\mathbf{k}}^{+}d\tau+N_{0}^{-1/2}\mathbf{B}^{\left(\alpha^{+}\right)}\mathbf{dW},\end{array}\label{eq:SDEgeneral}\end{equation}
where we have used a scaled time, $\tau=\kappa\sqrt{N_{0}}t$ and have also normalized the molecular field by its maximum (initial)
value, i.e. $\alpha=\beta/\sqrt{N_{0}}$. The deterministic part of the It\=o equations corresponds to the drift terms in the FPE, which if taken alone, are equivalent to the so-called `pairing mean-field theory' \cite{JackPRA2005,DavisPRA2008,Ogren2010}.
The stochastic part, in which $\mathbf{B}^{\left(\lambda\right)}$ are
row vectors with elements that are functions of the phase-space variables,
and where $\mathbf{dW}$ is a column-vector of real Wiener increments,
constitutes diffusion processes in the complex phase-space. This form of Eqs.~(\ref{eq:SDEgeneral})
shows that with drift terms of order $1$, the stochastic terms are
of order $1/\sqrt{N_{0}}$, i.e. the stochastic terms and therefore
non-mean-field corrections are more important for
decreasing $N_{0}$. 

Stochastically sampled moments can be related to physical expectation values.  For example, the first order moments give:
\begin{equation}
\left\{ \begin{array}{lll}
\langle n_{\mathbf{k}}\rangle _{S} &=& \langle \widehat{n}_{\mathbf{k}}\rangle=\langle \widehat{b}_{\mathbf{k}}^{\dagger}\widehat{b}_{\mathbf{k}}\rangle,\\
\langle m_{\mathbf{k}} \rangle _{S} &=& \langle \widehat{m}_{\mathbf{k}}\rangle=\langle \widehat{b}_{\mathbf{k},1}\widehat{b}_{-\mathbf{k},2}\rangle,\\
\langle \alpha \rangle _{S} &=&\langle \widehat{a} \rangle / \sqrt{N_0}.\end{array}\right.\label{eq:FirstOrderSmoments}\end{equation}
Normally ordered higher-order moments are obtained
exactly by stochastic averages of a corresponding Wick decomposition~\cite{CorneyPRL2004},
as in the following example
\begin{equation}
\langle m_{\mathbf{k}}^{+}m_{\mathbf{k}}\rangle_{S}+\langle n_{\mathbf{k}}^{2}\rangle_{S} = \langle\hat{m}_{\mathbf{k}}^{\dagger}\hat{m}_{\mathbf{k}}\rangle.\label{eq:examplehigherordermoment}\end{equation}
Note, however, that this
does not mean that a Wick factorisation is assumed to hold for a general quantum
state~\cite{Ogren2010}, since the average of a product is not the same as the product of averages.

The equivalences above hold so long as the appropriate moments of the distribution are well-defined.  In practice this requires that the distribution tails vanish sufficiently quickly, which again places a limit on the simulation time, indicated by `spiking' behaviour and associated rapid growth of sampling error.  The instabilities underlying this behaviour are a general feature of nonlinear stochastic equations~\cite{GilchristPRA1997, Astrom1965}.

With the equivalences between stochastic averages and operator expectation values,
the defined conserved quantities $F_\mathbf{k}$ [Eq.~(\ref{MPMQaveragerelation})], $E$ and $N$ can be calculated:
\begin{gather}
F_{\mathbf{k}}=\langle m_{\mathbf{k}}^{+}m_{\mathbf{k}}+n_{\mathbf{k}%
}^{2}-n_{\mathbf{k}}\rangle _{S},  \label{eq:testfunction} \\
E=\frac{1}{|\Delta |}\sum_{\mathbf{k}}\langle \Delta _{\mathbf{k}}n_{\mathbf{%
k}}-i\frac{\kappa \sqrt{N_{0}}}{2}\left( \alpha ^{+}m_{\mathbf{k}}-\alpha m_{%
\mathbf{k}}^{+}\right) \rangle _{S},  \label{eq:Es} \\
N=\left\langle \alpha ^{+}\alpha \right\rangle _{S}+\frac{1}{N_{0}}\sum_{%
\mathbf{k}}\left\langle n_{\mathbf{k}}\right\rangle _{S}.  \label{eq:Ns}
\end{gather}%
Note that although the stochastic quantities defined in Eqs.~(\ref{eq:testfunction})-(\ref{eq:Ns}) are complex for individual trajectories,
the average of the imaginary components approach zero with increasingly
many stochastic trajectories sampled.    Thus the average of each of these quantities approaches a real value, as expected for physical observables.

\begin{figure}

\includegraphics[scale=0.32]{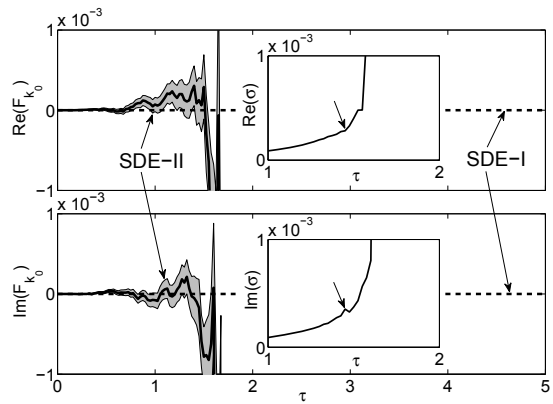}

\caption{Real and imaginary part of the conserved quantity $F_{\mathbf{k}_0}$, defined in Eq.~(\ref{eq:testfunction}), as a function of scaled time $\tau$.
Dashed line gives the mean for SDE-I, while the sampling error vanishes for SDE-I (see text).
Solid line gives the mean for SDE-II, with the light-grey shading giving the sampling error ($\pm \sigma$).
The spiking behaviour and rapid growth of sampling error  at $\tau = 1.5$ mean that the results from SDE-II cannot be used past this time.
The insets show the standard deviation $\sigma$ of $F_{\mathbf{k}_0}$ for SDE-II, with the arrows pointing out precursors of the spiking behaviour (see text).
We use a momentum grid with $|k|=0, \: dk, \: ..., \: 1000 dk, \: \: dk=2\pi/L \simeq 7.92\times 10^3$ m$^{-1}$ and a resonance momenta $k_0= \sqrt{2m_a |\Delta|/\hbar}=500dk$, where $\Delta=-12500$ s$^{-1}$.
Initially we have $N_0=100$ molecules and an atomic vacuum. The atom-molecule coupling strength is $\kappa \simeq 500$ s$^{-1}$.
The stochastic quantities for both SDE-I and SDE-II are evaluated using $10^4$ trajectories.
}

\label{fig1}
\end{figure}


The stochastic equations corresponding to a given Hamiltonian are not unique and therefore can be tailored to give different numerical and sampling properties~\cite{GardinerBook1}.  We illustrate how this can be done through the choice of `diffusion gauges' to extend the useful simulation time~\cite{CorneyPRL2004,PlimakPRA2001,DeuarPRA2002}.
The stochastic terms must fulfill the matrix-square-root condition~\cite{GardinerBook1} that relates the diffusion matrix $D$ in the Fokker-Planck equation 
to the noise-matrix $B$ :

\begin{equation}
D=BB^{T},\: B=\left[\begin{array}{c}
\mathbf{B}^{\left(n_{\mathbf{k}}\right)},
\mathbf{B}^{\left(m_{\mathbf{k}}\right)},
\mathbf{B}^{\left(m_{\mathbf{k}}^{+}\right)},
\mathbf{B}^{\left(\alpha\right)},
\mathbf{B}^{\left(\alpha^{+}\right)} \end{array}\right]^T,\label{eq:matrix-square-root-condition}\end{equation}
where $T$ denotes matrix transpose.  Let $O$ denote a matrix with orthonormal rows composed of functions of phase-space variables. Then if $B$ fulfills Eq.~(\ref{eq:matrix-square-root-condition}),  so does $\tilde{B}=BO$, which gives infinitely many choices of the SDE.

One specific noise matrix, which we together with Eq.~(\ref{eq:SDEgeneral}) label SDE-I, is
\begin{equation}
B_{I}=\left[\begin{array}{cccc}
n_{\mathbf{k}}m_{\mathbf{k}} & -in_{\mathbf{k}}m_{\mathbf{k}} & n_{\mathbf{k}}m_{\mathbf{k}}^{+} & -in_{\mathbf{k}}m_{\mathbf{k}}^{+}\\
m_{\mathbf{k}}^{2} & -im_{\mathbf{k}}^{2} & -n_{\mathbf{k}}^{2} & in_{\mathbf{k}}^{2}\\
-n_{\mathbf{k}}^{2} & in_{\mathbf{k}}^{2} & m_{\mathbf{k}}^{+2} & -im_{\mathbf{k}}^{+2}\\
1 & i & 0 & 0\\
0 & 0 & 1 & i\end{array}\right],\label{SDE-I}
\end{equation}
where $\mathbf{dW}_{I}=  \left[\begin{array}{cccc}
dw_{1} & dw_{2} & dw_{3} & dw_{4}\end{array}\right]^{T}/\sqrt{2}$. Note that it is often notationally convenient to work instead with
complex Wiener increments, e.g. $dW^{\left(1\right)}=\left(dw_{1}+idw_{2}\right)/\sqrt{2}$,
such that $dW^{\left(j\right)}$ satisfies $\langle dW^{\left(j\right)}\left(\tau\right)dW^{\left(j'\right)}\left(\tau^{\prime}\right)\rangle=0,\,\langle dW^{\left(j\right)}\left(\tau\right)dW^{\left(j'\right)\ast}\left(\tau^{\prime}\right)\rangle=\delta_{jj^{\prime}}\delta\left(\tau-\tau^{\prime}\right)d\tau$.  Then, for example,  $\mathbf{B}_{I}^{\left(n_{\mathbf{k}}\right)}\mathbf{dW}_{I}=n_{\mathbf{k}}\left(m_{\mathbf{k}}dW^{\left(1\right)*}+m_{\mathbf{k}}^{+}dW^{\left(2\right)*}\right)$.

\begin{figure}
\includegraphics[scale=0.37]{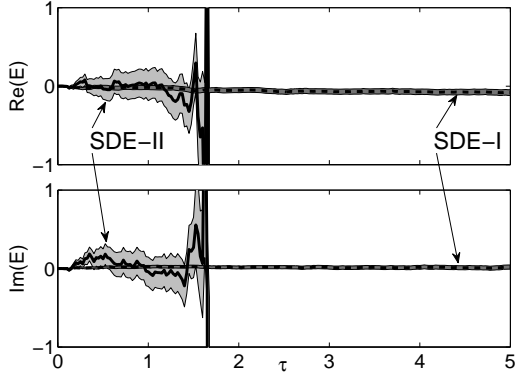}\caption{Normalised total energy $E$, defined in Eq.~(\ref{eq:Es}), as a function of scaled time $\tau$.
Dashed line and dark-grey shading give the mean and sampling error ($\pm \sigma$) for SDE-I.
Solid line and light-grey shading give the mean and sampling error ($\pm \sigma$) for SDE-II.
While the sampling error grows in both cases, SDE-I is stable for at least 3 times longer.
Parameters are as in Fig. \ref{fig1}.}
\label{fig2}
\end{figure}

\begin{figure}
\includegraphics[scale=0.37]{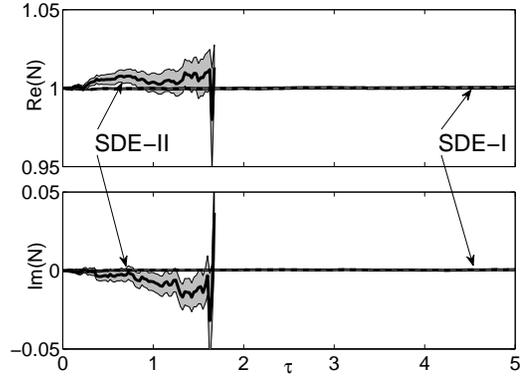}\caption{Normalised total particle number as defined in Eq.~(\ref{eq:Ns}), as a function of scaled time $\tau$.
Dashed line and dark-grey shading give the mean and sampling error ($\pm \sigma$) for SDE-I.
Solid line and light-grey shading give the mean and sampling error ($\pm \sigma$) for SDE-II.
Parameters are as in Fig. \ref{fig1}.}
\label{fig3}
\end{figure}

As proved in Appendix A, this choice of noise terms means that the quantity $F_{\mathbf{k}}$ defined in Eq.~(\ref{eq:testfunction}) is satisfied by each individual trajectory, not just by the ensemble average, i.e.
\begin{equation}
m_{\mathbf{k}}^{+}m_{\mathbf{k}}+n_{\mathbf{k}}^{2}=n_{\mathbf{k}}.\label{eq:ITOequality}
\end{equation}
This property is clearly seen graphically in Fig. \ref{fig1} as a vanishing sampling error for SDE-I.
In Figs. \ref{fig2} and \ref{fig3}, we see that the energy and particle number are conserved for SDE-I, but with a finite sampling error (dark-grey shading) which can be reduced further by including more stochastic trajectories.
The trajectories are stable, with no `spiking' or dramatic increase in sampling error, until at least a normalised time of $\tau = 5.0$.
We conclude that SDE-I performs very well for the particular set of parameters chosen.

We now use the gauge freedom of the condition in Eq.~(\ref{eq:matrix-square-root-condition})
to construct another specific noise matrix (SDE-II) which does not
fulfil Eq.~(\ref{eq:ITOequality}). For the case of a single $\bf k$-mode, the noise matrix for this diffusion gauge can be written:
\begin{equation}
B_{II}= \left[\begin{array}{cccccccc}
n_{\mathbf{k}} & -in_{\mathbf{k}} & n_{\mathbf{k}} & -in_{\mathbf{k}} & 0 & 0 & 0 & 0\\
m_{\mathbf{k}} & -im_{\mathbf{k}} & 0 & 0 & 0 & 0 & -n_{\mathbf{k}} & in_{\mathbf{k}}\\
0 & 0 & m_{\mathbf{k}}^{+} & -im_{\mathbf{k}}^{+} & -n_{\mathbf{k}} & in_{\mathbf{k}} & 0 & 0\\
m_{\mathbf{k}} & im_{\mathbf{k}} & 0 & 0 & n_{\mathbf{k}} & in_{\mathbf{k}} & 0 & 0\\
0 & 0 & m_{\mathbf{k}}^{+} & im_{\mathbf{k}}^{+} & 0 & 0 & n_{\mathbf{k}} & in_{\mathbf{k}}\end{array}\right].\label{SDE-II-1}
\end{equation}
However, in general $B_{II}$ is of size $\left(3M+2\right)\times8M$,
i.e.  the number of noise columns grows with the number of phase-space
variables, such that now $\mathbf{dW}_{II}=  \left[\begin{array}{cccc}
dw_{1,\mathbf{k}} & dw_{2,\mathbf{k}} & ... & dw_{8,\mathbf{k}}\end{array}\right]^{T} /\sqrt{2} $. In this case we have for example
$\mathbf{B}_{II}^{\left(n_{\mathbf{k}}\right)}\mathbf{dW}_{II}=n_{\mathbf{k}}\left(dW_{\mathbf{k}}^{\left(1\right)*}+dW_{\mathbf{k}}^{\left(2\right)*}\right)$
and $\mathbf{B}_{II}^{\left(\alpha\right)}\mathbf{dW}_{II}=\sum_{\mathbf{k}}m_{\mathbf{k}}\left(dW_{\mathbf{k}}^{\left(1\right)}+dW_{\mathbf{k}}^{\left(3\right)}\right)$.

For SDE-II it is now only the average of $F_{\mathbf{k}}$ that is zero, within the finite sampling error indicated by the light-grey shading in Fig. \ref{fig1}.  For the energy and particle number, the average is still constant within the sampling error,  as  shown in Figs. \ref{fig2} and \ref{fig3}.  However, the sampling error is now larger than for SDE-I.  Moreover, the mean results from SDE-II (solid line in all figures) start to spike before $\tau\sim1.5$, with an associated dramatic increase in sampling error, and can thus not be used beyond this point for the present parameters.
The standard deviation of a stochastic variable is a moment of higher order than the average of the variable itself, and precursors of the spiking behaviour are first seen here.
This is illustrated by the arrows in the inset plots of Fig. \ref{fig1} for the standard deviation of $F_{\mathbf{k}_0}$, but generally occurs for all sampled variables.

As shown in Figs.~\ref{fig1}-\ref{fig3}, there can be dramatic differences in the performance of the two diffusion
gauges.  However the relative performance depends on the system parameters.
For instance, whereas the second gauge (SDE-II) may seem unnecessarily complicated for the many modes, and leads to a much larger sampling error and a shorter useful simulation time here,
for the small system in \cite{Ogren2010}, it is in fact superior to SDE-I in terms of useful simulation time.


From the theoretical foundation it is expected that
the Gaussian phase-space method is exact while the distribution is sufficiently bounded \cite{CorneyPRL2004}. In practice
the simulation can be trusted until signatures
such as spiking trajectories and rapid growth of the sampling error occurs in the time evolution of the phase-space variables \cite{GilchristPRA1997, DeuarPRA2002, CorbozPRB2008, Ogren2010}.
We have previously also analysed a related dynamical system with only $N_{0}=10$ molecules
and $M=10$ atomic momentum modes \cite{Ogren2010}. For this test system, the
exponentially growing dimension of the Hilbert space was small enough
($d=2^{M}n_{\mathrm{max}}\simeq10^{5}$), to allow a direct comparison to an expansion in a number
state basis. 
However, this comparison is not possible for the system under study here.
Having explicit access to different stochastic realisations of the FPE, as here with Eqs.~(\ref{SDE-I}) and
(\ref{SDE-II-1}), then gives the possibility to compare different stochastic calculations of the moments to check the accuracy of the numerical implementation
or to detect errors in the underlying derivations.


Despite the different stochastic behaviour revealed in Figs.~\ref{fig1}-\ref{fig3}, it is important to note that SDE-I and SDE-II
both correspond to the same Hamiltonian (\ref{Hamiltonian}) and the same complex FPE Eq.~(\ref{PDE}). Underlying these different realisations is the overcompleteness of the Gaussian representation, which allows the one density operator $\hat \rho$ to be mapped to many different distributions.

In summary, we have demonstrated how different diffusion gauges can substantially change the numerical performance of the Gaussian fermionic phase-space method.  This ability to manipulate the form of stochastic equation can be used to reduce the sampling error and extend the useful simulation time, depending on the system parameters.
In addition, we have shown that the simulation of conserved quantities can have qualitatively different behaviour for different gauges.  The conserved quantities thus provide a check on numerical implementation and allow the performance of different gauges to be benchmarked.

\section{Acknowledgments}

The authors acknowledge support by the Australian Research Council.
We would also like to thank the developers of the \emph{xmds} software \cite{xmds} used in our simulations.
M.\"{O}. especially thanks G. Dennis and J. Hope for
valuable advice during a research visit at the Australian National
University and the Solander program at the University of Queensland for financial support.


\section{Appendix A. Derivation of Eq. (\ref{eq:ITOequality})   } 

Here we prove Eq.~(\ref{eq:ITOequality}) for SDE-I, which is a
stronger condition than the corresponding result for the stochastic average. We apply the product rule for two stochastic
variables $X$ and $Y$ within Ito calculus
\begin{equation}
d_{I}\left(XY\right)=Xd_{I}\left(Y\right)+d_{I}\left(X\right)Y+d_{I}\left(X\right)d_{I}\left(Y\right) \label{ItoRule},
\end{equation}
to the first term in Eq.~(\ref{eq:ITOequality}), with $d_{I}$ denoting the Ito differential.
Hence we have, from Eqs.~(\ref{eq:SDEgeneral}) and~(\ref{SDE-I})
\begin{equation}
\begin{array}{l}
d_{I}\left(m_{\mathbf{k}}^{+}m_{\mathbf{k}}\right)=-2i\delta_{\mathbf{k}}m_{\mathbf{k}}^{+}m_{\mathbf{k}}d\tau+\alpha m_{\mathbf{k}}^{+}\left(1-2n_{\mathbf{k}}\right)d\tau\\
+N_{0}^{-1/2}m_{\mathbf{k}}^{+}\left(m_{\mathbf{k}}^{2}dW_{1}^{\ast}-n_{\mathbf{k}}^{2}dW_{2}^{\ast}\right)+2i\delta_{\mathbf{k}}m_{\mathbf{k}}^{+}m_{\mathbf{k}}d\tau\\
+\alpha^{+}\left(1-2n_{\mathbf{k}}\right)m_{\mathbf{k}}d\tau+N_{0}^{-1/2}\left(m_{\mathbf{k}}^{+2}dW_{2}^{\ast}-n_{\mathbf{k}}^{2}dW_{1}^{\ast}\right)m_{\mathbf{k}}\\
+N_{0}^{-1}\left(m_{\mathbf{k}}^{+2}dW_{2}^{\ast}-n_{\mathbf{k}}^{2}dW_{1}^{\ast}\right)\left(m_{\mathbf{k}}^{2}dW_{1}^{\ast}-n_{\mathbf{k}}^{2}dW_{2}^{\ast}\right)\\
=\left(\alpha m_{\mathbf{k}}^{+}+\alpha^{+}m_{\mathbf{k}}\right)\left(1-2n_{\mathbf{k}}\right)d\tau\\
+N_{0}^{-1/2}\left(m_{\mathbf{k}}^{+}m_{\mathbf{k}}-n_{\mathbf{k}}^{2}\right)\left(m_{\mathbf{k}}dW_{1}^{\ast}+m_{\mathbf{k}}^{+}dW_{2}^{\ast}\right),\end{array} \label{dummylabel}
\end{equation}
where we have kept, as usual, only first order terms in $d\tau$.  The increment for $n_{\mathbf{k}}^{2}$ can be calculated similarly,
leading to the following equation for the increment in the left-hand side of Eq.~(\ref{eq:ITOequality}):
\begin{equation}
\begin{array}{l}
d_{I}\left(m_{\mathbf{k}}^{+}m_{\mathbf{k}}+n_{\mathbf{k}}^{2}\right)=\left(\alpha m_{\mathbf{k}}^{+}+\alpha^{+}m_{\mathbf{k}}\right)d\tau\\
+N_{0}^{-1/2}\left(m_{\mathbf{k}}^{+}m_{\mathbf{k}}+n_{\mathbf{k}}^{2}\right)\left(m_{\mathbf{k}}dW_{1}^{\ast}+m_{\mathbf{k}}^{+}dW_{2}^{\ast}\right).\end{array}\label{eq:resultOfDerivationAppA}\end{equation}
From Eqs.~(\ref{eq:SDEgeneral}) and (\ref{SDE-I}), the corresponding expression for the left-hand side of Eq.~(\ref{eq:ITOequality}) is
\begin{equation}
dn_{\mathbf{k}}=\left(\alpha m_{\mathbf{k}}^{+}+\alpha^{+}m_{\mathbf{k}}\right)d\tau + N_{0}^{-1/2} n_{\mathbf{k}} \left(m_{\mathbf{k}}dW_{1}^{\ast}+m_{\mathbf{k}}^{+}dW_{2}^{\ast}\right).\label{RHS}
\end{equation}
The initial conditions are $m_{\mathbf{k}}^{+}=m_{\mathbf{k}}=n_{\mathbf{k}}=0$, which satisfy the equality  (\ref{eq:ITOequality}) trivially.  If initially true, then Eqs.\ (\ref{eq:resultOfDerivationAppA}) and (\ref{RHS})  guarantee the equality for consecutive time-steps of SDE-I.

However, it is straightforward
to show that any stochastic gauge that does not have the same indices on the noises for  $d_{I}\left(n_{\mathbf{k}}\right)$ and
$d_{I}\left(m_{\mathbf{k}}\right)$ does not fulfill Eq.~(\ref{eq:ITOequality}). This is in particular exemplified with SDE-II and the qualitative difference in the sampling errors of $F_{\mathbf{k}}$ for the two gauges is seen in Fig. \ref{fig1}.


\begin{thebibliography}{29}
\bibitem{DeuarPRL2007} P.~Deuar and P.~D.~Drummond, Phys. Rev.
Lett. \textbf{98}, 120402 (2007); A.~Perrin \emph{et al.}, New J.
Physics \textbf{10}, 045021 (2008).

\bibitem{SavagePRA2006} C.~M.~Savage, P.~E.~Schwenn, and K.~V.~Kheruntsyan,
Phys. Rev. A \textbf{74}, 033620 (2006).

\bibitem{Linden1992} W.~von der Linden, Physics Reports \textbf{220},
53 (1992).

\bibitem{MakJCP2009} C.~H.~Mak, J. Chem. Phys. \textbf{131}, 044125
(2009).

\bibitem{MontinaPRA2006} O.~Juillet, F.~Gulminelli, and Ph.~Chomaz,
Phys. Rev. Lett. \textbf{92} 160401 (2004); A.~Montina and Y.~Castin,
Phys. Rev. A \textbf{73}, 013618 (2006). 


\bibitem{CorneyPRL2004} J.~F.~Corney and P. D. Drummond, Phys.
Rev. Lett. \textbf{93}, 260401 (2004); Phys. Rev. B \textbf{73}, 125112
(2006); J. Phys. A: Math. Gen. \textbf{39}, 269 (2006).

\bibitem{JackPRA2005} M.~W.~Jack and H.~Pu, Phys. Rev. A \textbf{72},
063625 (2005).

\bibitem{KheruntsyanPRL2006} K.~V.~Kheruntsyan, Phys. Rev. Lett.
\textbf{96}, 110401 (2006).

\bibitem{FriedbergPRB1989} R.~Friedberg and T.~D.~Lee, Phys. Rev.
B \textbf{40}, 6745 (1989).


\bibitem{Ogren2010}M.~\"{O}gren, K.~V.~Kheruntsyan, and J.~F.~Corney, 
Europhys. Lett. \textbf{92} 36003 (2010).

\bibitem{GilchristPRA1997} A.~Gilchrist, C.~W.~Gardiner, and P.~D.~Drummond,
Phys. Rev. A \textbf{55} 3014, (1997).

\bibitem{PlimakEPL2001}L. I. Plimak \emph{et al.}, Europhys.
Lett. \textbf{56} 372 (2001).

\bibitem{RahavPRB2009} S.~Rahav and S.~Mukamel, Phys. Rev. B \textbf{79},
165103 (2009).

\bibitem{GardinerBook1} C.~W.~Gardiner, \textit{Handbook of Stochastic
Methods}, Springer, 4th ed. (Springer, Berlin, 2008).

\bibitem{NumSim} We convert the Ito equations to Stratonovich form and
integrate with a semi-implicit method \cite{DrummondJComputPhys1991}
that has better convergence properties.
For SDE-II, this means that non-zero Stratonovich corrections to the drift,
of the form $-\sum_{j,k}B_{j,k}\partial_{j}B_{i,k}/2$, need to be added \cite{GardinerBook1},
whereas for SDE-I, those corrections were zero.

\bibitem{DrummondJComputPhys1991}P. D. Drummond and I. K. Mortimer,
J. Comput. Phys. \textbf{93}, 144 (1991).

\bibitem{DavisPRA2008} M.~J.~Davis \emph{et al.}, Phys. Rev. A
\textbf{77}, 023617 (2008).

\bibitem{Astrom1965}K. J. \AA str\"{o}m, International Journal of Control
\textbf{1}, 301 (1965).

\bibitem{PlimakPRA2001}L. I. Plimak, M. K. Olsen, and M. J. Collett,
Phys. Rev. A \textbf{64} 025801, (2001).

\bibitem{DeuarPRA2002}P. Deuar and P. D. Drummond, Phys. Rev. A \textbf{66}
033812, (2002).

\bibitem{CorbozPRB2008} P.~Corboz \emph{et al.}, Phys. Rev. B \textbf{77},
085108 (2008).

\bibitem{xmds} \textit{www.xmds.org}

\end{thebibliography}
\end{document}